\documentclass[aps,prd,preprint,groupedaddress,amsmath,amssymb]{revtex4}
\usepackage{slashed}
\usepackage{epsfig}
\usepackage{multirow}

\begin{document}

\def\Re  {{\rm Re}}
\def\Im  {{\rm Im}}
\def\KFS#1{K^{{#1}}_{\scriptscriptstyle\rm F\!.S\!.}}
\def\Ret {\widetilde{{\rm Re}}}
\def\veps{\varepsilon}
\def\hc    {{\rm h.c.}}
\renewcommand{\eqref}[1]{Eq.~(\ref{#1})}
\newcommand{\fact}{{\mathrm{C}}}
\newcommand{\figref}[1]{Fig.~\ref{#1}}
\newcommand{\tabref}[1]{Tab.~\ref{#1}}
\newcommand{\rA}{{\mathrm{A}}}
\newcommand{\rB}{{\mathrm{B}}}
\newcommand{\rR}{{\mathrm{R}}}
\newcommand{\rV}{{\mathrm{V}}}
\def\bom#1{{\mbox{\boldmath $#1$}}}
\def\ket#1{|{#1}\rangle}
\def\bra#1{\langle{#1}|}
\def\sd{\tilde{d}}
\def\sq{\tilde{q}}
\def\su{\tilde{u}}
\def\mq{m_q}
\def\mqi{m_{q}}
\def\mqj{m_{q}}
\def\mqk{m_{q}}
\def\mql{m_{q}}
\def\msqi{m_{\tilde{q}_i}}
\def\msqj{m_{\tilde{q}_j}}
\def\msqk{m_{\tilde{q}_k}}
\def\msql{m_{\tilde{q}_l}}
\def\mgl{m_{\tilde{g}}}

\def\as{\alpha_s}
\def\nbar{\bar{N}}
\def\bbar{\bar{b}}

\def\ca{\tilde{\chi}^\pm_1}
\def\cb{\tilde{\chi}^\pm_2}
\def\cc{\tilde{\chi}^-_1}
\def\cd{\tilde{\chi}^-_2}
\def\ce{\tilde{\chi}^+_2}
\def\cf{\tilde{\chi}^+_1}
\def\na{\tilde{\chi}^0_1}
\def\nb{\tilde{\chi}^0_2}
\def\nc{\tilde{\chi}^0_3}
\def\nd{\tilde{\chi}^0_4}
\def\ncd{\tilde{\chi}^0_{3,4}}
\def\sq{\tilde{q}}

\def\md{m_{\tilde{\chi}_1^0}}
\def\mn{m_{\tilde{\chi}_2^0}}
\def\mc{m_{\tilde{\chi}_1^\pm}}
\def\ms{m_{\tilde{q}}}
\def\ml{m_{\tilde{l}}}

\def\cA{{\cal A}} \def\cB{{\cal B}} \def\cC{{\cal C}} \def\cD{{\cal D}}
\def\cE{{\cal E}} \def\cF{{\cal F}} \def\cG{{\cal G}} \def\cH{{\cal H}}
\def\cI{{\cal I}} \def\cJ{{\cal J}} \def\cK{{\cal K}} \def\cL{{\cal L}}
\def\cM{{\cal M}} \def\cN{{\cal N}} \def\cO{{\cal O}} \def\cP{{\cal P}}
\def\cQ{{\cal Q}} \def\cR{{\cal R}} \def\cS{{\cal S}} \def\cT{{\cal T}}
\def\cU{{\cal U}} \def\cV{{\cal V}} \def\cW{{\cal W}} \def\cX{{\cal X}}
\def\cY{{\cal Y}} \def\cZ{{\cal Z}} 

\def\d{{\rm d}}
\def\eps{\epsilon}
\def\MSbar{\overline{\rm MS}}

\def\lp{\left. }
\def\rp{\right. }
\def\lr{\left( }
\def\rr{\right) }
\def\le{\left[ }
\def\re{\right] }
\def\lg{\left\{ }
\def\rg{\right\} }
\def\lb{\left| }
\def\rb{\right| }

\def\bsp#1\esp{\begin{split}#1\end{split}}

\def\beq{\begin{equation}}
\def\eeq{\end{equation}}
\def\bea{\begin{eqnarray}}
\def\eea{\end{eqnarray}}

\preprint{IPHC-PHENO-11-01}
\preprint{LPSC 11-031}
\preprint{MS-TP-11-02}
\title{Joint Resummation for Gaugino Pair Production \\ at Hadron
 Colliders}
\author{Jonathan Debove$^a$}
\author{Benjamin Fuks$^b$}
\author{Michael Klasen$^{a,c}$}
\email[]{klasen@lpsc.in2p3.fr}
\affiliation{$^a$ Laboratoire de Physique Subatomique et de Cosmologie,
 Universit\'e Joseph Fourier/CNRS-IN2P3/INPG,
 53 Avenue des Martyrs, F-38026 Grenoble, France \\
 $^b$ Institut Pluridisciplinaire Hubert Curien/D\'epartement Recherche
 Subatomique, Universit\'e de Strasbourg/CNRS-IN2P3, 23 Rue du Loess,
 F-67037 Strasbourg, France \\
 $^c$ Institut f\"ur Theoretische Physik, Westf\"alische
 Wilhelms-Universit\"at M\"unster, Wilhelm-Klemm-Stra\ss{}e 9, D-48149 M\"unster,
 Germany}
\date{\today}
\begin{abstract}
 We calculate direct gaugino pair production at hadron colliders at
 next-to-leading order of perturbative QCD, resumming simultaneously large
 logarithms in the small transverse-momentum and threshold regions to
 next-to-leading logarithmic accuracy. Numerical predictions are presented for
 transverse momentum and invariant mass spectra as well as for total cross
 sections and compared to results obtained at fixed order and with pure
 transverse-momentum and threshold resummation. We find that our new results are
 in general in good agreement with the previous ones, but often even more precise.
\end{abstract}
\maketitle



\section{Introduction}
\label{sec:1}

Weak-scale supersymmetry (SUSY), and in particular the Minimal Supersymmetric
Standard Model (MSSM), is a theoretically and phenomenologically well-motivated
extension of the Standard Model (SM) of particle physics \cite{Nilles:1983ge}.
Consequently, the experimental search for the spin partners predicted by the MSSM
for each of the SM particles is one of the defining tasks at
current high-energy colliders such as the $p\bar{p}$ Tevatron collider at Fermilab
\cite{Aaltonen:2008pv} and the $pp$ Large Hadron Collider (LHC) at CERN
\cite{Aad:2009wy,Ball:2007zza}. Of particular interest are the neutral and charged
fermionic
partners of the electroweak gauge and Higgs bosons, which mix due to their equal
quantum numbers into four neutralino $(\tilde{\chi}^0_i$, $i=1,\dots,4$) and
chargino $(\tilde{\chi}^\pm_i$, $i=1,2$) mass eigenstates, since these participate
virtually always in SUSY collider signatures and may also have
important implications for dark matter and cosmology. Their decays into leptons
and missing transverse energy, carried away by the lightest SUSY particle (LSP,
often the $\na$), are easily identifiable at hadron colliders. The lighter
gaugino/higgsino mass eigenstates are accessible not only at the LHC with
center-of-mass energies $\sqrt{S}$ of 7 to 14 TeV \cite{Aad:2009wy,Ball:2007zza},
but also at Run II of the Tevatron ($\sqrt{S} =1.96$ TeV), where the production of
$\ca\nb$ pairs decaying into trilepton final states is one of the gold-plated SUSY
discovery channels \cite{Aaltonen:2008pv}.

For an efficient suppression of the SM background from vector-boson and top-quark
production and a precise determination of the underlying SUSY-breaking model and
masses, an accurate theoretical calculation of the signal (and background) cross
section is imperative. As the LSP escapes undetected, the key distribution for
SUSY discovery and measurements is the missing transverse-energy ($\not{\!\!E}_T$)
spectrum, which is typically restricted in the experimental analyses by a cut of
20 GeV at the Tevatron and 30 GeV at the LHC. While the SUSY particle pair is
produced with zero transverse
momentum ($p_T$) in the Born approximation, the possible radiation of gluons from
the quark-antiquark initial state or the splitting of gluons into quark-antiquark
pairs at ${\cal O}(\alpha_s)$ in the strong coupling constant induces transverse
momenta extending to quite substantial values and must therefore be taken into
account. In addition, the perturbative calculation diverges at small $p_T$,
indicating the need for a resummation of soft-gluon radiation to all orders.
Only after a consistent matching of the perturbative and resummed calculations an
accurate description of the (missing) transverse energy spectrum and precise
measurements of the SUSY particle masses can be achieved. Furthermore, when the
SUSY particle pair with invariant mass $M$ is produced close to the production
threshold at the partonic center-of-mass energy $s$, soft gluon emission leads
again to potentially large (logarithmic) terms, which must be resummed to all
orders in order to obtain a reliable cross section.
The production of SUSY particles at hadron colliders has been studied at leading
order (LO) of perturbative QCD since the 1980s \cite{Barger:1983wc}. More
recently, previously neglected electroweak contributions \cite{Bozzi:2005sy},
polarization effects \cite{Gehrmann:2004xu}, and the violation of flavor
\cite{Bozzi:2007me} and $CP$ symmetry \cite{Alan:2007rp} have been considered at
this order. Next-to-leading order (NLO) corrections have been computed within QCD
since the late 1990s \cite{Beenakker:1996ch} and recently also
within the electroweak theory \cite{Hollik:2007wf}. Resummation at the
next-to-leading logarithmic (NLL) level has been achieved in the small-$p_T$
region for sleptons and gauginos \cite{Bozzi:2006fw,Debove:2009ia} and in the
threshold region for sleptons, gauginos, squarks and gluinos \cite{Bozzi:2007qr,%
Debove:2010kf}.

Since the dynamical origin of the enhanced contributions is the same both in
transverse-momentum and threshold resummations, {\em i.e.} the soft-gluon emission
by the initial state, it would be desirable to have a formalism capable to handle
at the same time the soft-gluon contributions in both the delicate kinematical
regions, $p_{T} \ll M$ and $M^2\sim s$. This {\it joint} resummation formalism
has been developed over the last twelve years \cite{Li:1998is}. The exponentiation
of the singular terms in the Mellin ($N$) and impact-parameter ($b$) space has
been proven, and a consistent method to perform the inverse transforms, avoiding
the Landau pole and the singularities of the parton distribution functions (PDFs),
has been introduced \cite{Laenen:2000de}. Applications to prompt-photon,
electroweak and Higgs boson, heavy-quark and slepton pair production at
hadron colliders have exhibited substantial effects of joint resummation on the
differential and total cross sections \cite{Laenen:2000ij}.

In this paper, we present the first calculation of joint resummation of soft
gluon effects in the small transverse momentum and threshold regions for
gaugino/higgsino hadroproduction at the NLL level, using the formalism described
above. As in our previous calculations \cite{Bozzi:2006fw,Bozzi:2007qr,%
Debove:2009ia,Debove:2010kf}, we include not only the QCD, but also the SUSY-QCD
virtual loop contributions with internal squark mixing in the hard coefficient
function of the resummed cross section, which therefore reproduces, when expanded
and integrated over $p_T$, the correct NLO SUSY-QCD cross section in the threshold
region. For the Tevatron, we consider not only the production of $\ca\nb$, but
also of $\nb\nb$ and $\ca\ca$ pairs. In particular the latter can have
significantly larger cross sections than
trilepton production due to the $s$-channel exchange of massless photons. For the
LHC, we concentrate on predictions for its current center-of-mass energy of
$\sqrt{S}=$7 TeV, where threshold effects and direct gaugino pair production (as
opposed to the production from squark and gluino cascade decays) will be more
important. However, we will also show cross sections for the production of light
($\ca$, $\nb$) and heavy ($\cb$, $\ncd$) gaugino combinations at the LHC design
energy of $\sqrt{S}=$14 TeV for completeness.

The remainder of this paper is organized as follows: In Sec.\ \ref{sec:2},
we briefly review the joint resummation formalism in Mellin and impact parameter
space, giving the explicit form of the resummed logarithms at NLL order as well as
their matching to the fixed order calculation and the prescriptions employed for
the inverse integral transforms. Sec.\ \ref{sec:3} contains numerical results for
various gaugino pair production cross sections at the Tevatron and at the LHC
and compares the results obtained in joint resummation to those obtained with
the pure $p_T$- and threshold resummation formalisms, respectively. We summarize
our results in Sec.\ \ref{sec:4}.

\section{Joint resummation formalism}
\label{sec:2}

Thanks to the QCD factorization theorem, the double differential cross section
for the hadronic production of a final state with fixed invariant mass $M$ and
transverse momentum $p_T$
\bea
  M^2\frac{\d^2\sigma_{AB}}{\d M^2 \d p_T^2}(\tau)&=&\sum_{ab}
  \int_0^1 \!\d x_a \d x_b \d z[x_a f_{a/A}(x_a,\mu_F^2)] [x_b f_{b/B}(x_b,
  \mu_F^2)]\,[z\,\d\sigma_{ab}(z,M^2,p_T^2,\mu_F^2)]\nonumber\\
 &\times& \delta(\tau-x_ax_bz)
  \label{eq:HadFacX}
\eea
can be obtained by convolving the partonic cross section $\d\sigma_{ab}$ with
the universal densities $f_{a,b/A,B}$ of the partons $a,b$, carrying the momentum
fractions $x_{a,b}$ of the colliding hadrons $A,B$, at the factorization scale
$\mu_F$. The application of a Mellin transform
\bea
  F(N)&=&\int_0^1 \d y \,y^{N-1} F(y)
  \label{eq:MelDef}
\eea
to the quantities $F\in\{\sigma_{AB},~\sigma_{ab},~f_{a/A},~f_{b/B}\}$ with
$y\in\{\tau=M^2/S,~z=M^2/s,~x_{a},~x_{b}\}$ allows to express the hadronic cross
section in moment space as a simple product,
\bea
  M^2\frac{\d^2\sigma_{AB}}{\d M^2 \d p_T^2}(N-1)&=&\sum_{ab} f_{a/A}(N,\mu_F^2) 
  f_{b/B}(N,\mu_F^2) \sigma_{ab}(N,M^2,p_T^2,\mu_F^2).
  \label{eq:HadFacN}
\eea
Furthermore, the application of a Fourier transform to the partonic cross section
$\sigma_{ab}$ allows to correctly take into account transverse-momentum
conservation, so that in moment ($N$) and impact parameter ($b$) space it can be
written as
\bea
 \sigma_{ab}(N,M^2,p_T^2,\mu_F^2)&=&
 \int_0^\infty \d b\frac{b}{2} J_0(bp_T) \sigma_{ab}(N,M^2,b^2,\mu_F^2).
 \label{eq:JR:pff}
\eea
Here, $J_0(y)$ denotes the $0^{\rm th}$-order Bessel function and
\bea
 \sigma_{ab}(N,M^2,b^2,\mu_F^2)&=&\sum_{n=0}^\infty a_s^{n}(\mu_R^2)\,
 \sigma_{ab}^{(n)}(N,M^2,b^2,\mu_F^2,\mu_R^2)
 \label{eq:5}
\eea
is usually expanded perturbatively in the strong coupling constant $a_s(\mu^2)=
\alpha_s(\mu^2)/(2\pi)$ at the renormalization scale $\mu_R$. For simplicity,
we identify in the following the factorization and renormalization scales, {\em
i.e.} $\mu_F=\mu_R=\mu$.

\begin{figure}
 \centering
 \epsfig{file=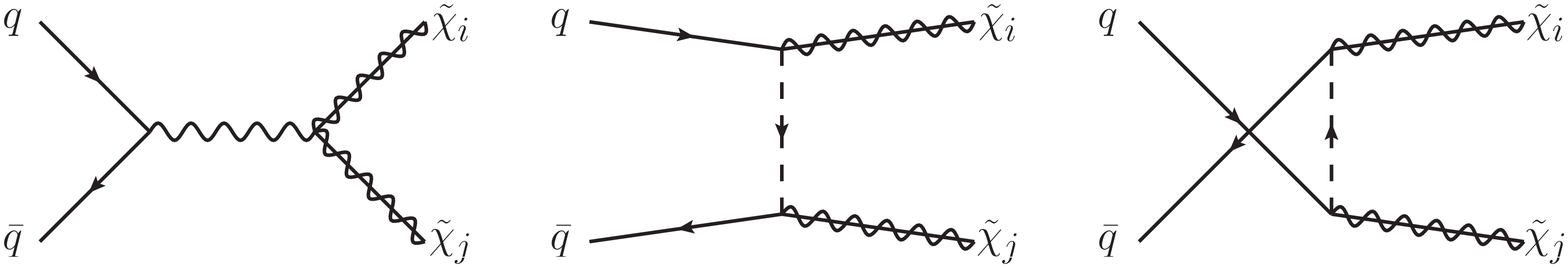,width=.75\columnwidth}
 \caption{\label{fig:0}Tree-level Feynman diagrams for the production of gaugino
          pairs.}
\end{figure}

In the Born approximation, the hadroproduction of neutralinos $\tilde{\chi}_i^0$
and charginos $\tilde{\chi}_i^\pm$ is induced by quarks $q$ and antiquarks
$\bar{q}'$ in the initial (anti-)protons and is mediated by $s$-channel
electroweak gauge-boson and $t$- and $u$-channel squark exchanges (see Fig.\
\ref{fig:0}). Its partonic cross section $\sigma_{q\bar{q}'}^{(0)}$ can be
%
%
expressed in terms of the gaugino and squark masses $m_{\tilde{\chi}^{0,\pm}
_{i,j}}$ and $m_{\tilde{q}}$, the masses of the electroweak gauge bosons, and
generalized charges \cite{Barger:1983wc}. At ${\cal O}(a_s)$, virtual loop and
real parton emission corrections must be taken into account
\cite{Beenakker:1996ch}. The latter induce not only a deviation of the partonic
center-of-mass energy $s$ from the squared invariant mass $M^2$ of the gaugino
pair, but also non-zero transverse momenta $p_T$, that extend typically to values
of the order of the gaugino mass.

Close to the partonic production threshold, where $z=M^2/s\to1$ or $N\to\infty$,
the convergence of the perturbative expansion is spoiled due to soft gluon
radiation, which induces large logarithms
\bea
 a_s^n\lr \ln^m(1-z)\over1-z\rr_+&\to&a_s^n\ln^{m+1}\nbar+\dots
\eea
with $m\leq2n-1$ and $\nbar=N e^{\gamma_E}$ \cite{Debove:2010kf}. Similarly, in
the small-$p_T$ (or large-$b$) region, where the bulk of the events is produced,
the convergence of the perturbative expansion is again spoiled by soft gluon
radiation, which induces large logarithms
\bea
 \alpha_s^n\lr{1\over p_T^2}\ln^m{M^2\over p_T^2}\rr_+&\to&
 \alpha_s^n\ln^{m+1}\bbar^2+\dots
\eea
with $m\leq2n-1$ and $\bbar=bMe^{\gamma_E}/2$ \cite{Debove:2009ia}. The crucial
observation, first
made by Li \cite{Li:1998is} and then further developped by Laenen, Sterman, and
Vogelsang \cite{Laenen:2000de,Laenen:2000ij} is that the common kinematic origin
of these divergences allows for a joint resummation of the large logarithms in the
partonic cross section by choosing a function
\bea
 \chi(\nbar, \bbar)&=&\frac{\nbar}{1+\eta\,\bbar/\nbar}+\bbar,
 \label{eq:chi}
\eea
which interpolates between $\nbar$ in the threshold region, $\nbar\gg\bbar$,
and $\bbar$ in the small-$p_T$ region, $\bbar\gg\nbar$. Its exact form is
constrained by the requirement that the leading and next-to-leading
logarithms in $\bbar$ and $\nbar$ are correctly reproduced in the limits
$\bar{b}\to \infty$ and $\bar{N}\to\infty$, respectively. The choice of Eq.\
(\ref{eq:chi}) with $\eta > 0$ (we use $\eta=1$) avoids the introduction of
sizeable subleading terms into perturbative expansions of the resummed cross
section at a given order in $a_s$, which are not present in fixed-order
calculations \cite{Laenen:2000ij}. Up to NLL order, the resummed cross section
can then be written in the form
\bea
 \sigma_{ab} (N,M^2,b^2,\mu^2)&=&\sum_{a',a'',b',b''}
 E_{a'a}^{(1)}(N,M^2/\chi^2,\mu^2) E_{b'b}^{(1)}(N,M^2/\chi^2,\mu^2)
 C_{a''a'}(N, a_s(M^2/\chi^2)) \nonumber\\
 &\times&
 C_{b''b'}(N, a_s(M^2/\chi^2))
 H_{a''b''}(M^2, \mu^2) \exp[G_{a''b''}(M^2,\bar{N},\bar{b}^2,\mu^2)],
 \label{eq:9}
\eea
which is very similar to the one of the $p_T$-resummed cross section. The
operators $E_{a'a,b'b}$
allow to evolve the PDFs $f_{a,b/A,B}(N,\mu^2)$ from the scale $\mu^2$ to the
scale $M^2/\chi^2$. They satisfy, like
the PDFs themselves, the Altarelli-Parisi equations \cite{Altarelli:1977zs} and
can be written to one-loop order, $E^{(1)}$, and in the singlet/non-singlet
basis in closed exponential form \cite{Furmanski:1981cw}. When the finite part of
the renormalized virtual one-loop contribution $\cA_0$, which in our calculation
contains the full SUSY-QCD corrections with internal squark mixing
\cite{Debove:2010kf}, is entirely absorbed into the hard function
\bea
  H_{ab}(M^2,\mu^2)&=&\sigma_{ab}^{(0)}(M^2,\mu^2)
  \big[1+a_s \cA_0\big]+\cO(a_s^2),
  \label{eq:PR:choice1}
\eea
the coefficients $C_{ab}$ become process-independent and can be written up to
one-loop order as $C_{ab}^{(0)}=\delta_{ab}$ and
\bea
  C_{ab}^{(1)}(N)&=&C_a\frac{\pi^2}{6}\delta_{ab}-P_{ab}^{(1),\eps}(N),
  \label{eq:PR:choice2}
\eea
where $C_{ab}=\sum_{n=0}^\infty a_s^{n}(\mu_R^2)\,C_{ab}^{(n)}$, the QCD color
factors are $C_q=C_F=4/3$ and $C_g=C_A=3$, respectively, and
\bea
  P_{qq}^{(1),\eps}(N)~=~\frac{-C_F}{N(N+1)}~~&,&~~
  P_{qg}^{(1),\eps}(N)~=~\frac{-2T_R}{(N+1)(N+2)},\\
  P_{gq}^{(1),\eps}(N)~=~\frac{-C_F}{N+1}\hspace*{8mm}&,&~~
  P_{gg}^{(1),\eps}(N)~=~0
  \label{eq:PR:EpsAP}
\eea
represent the $\cO(a_s,\eps)$ terms in the Altarelli-Parisi splitting functions
in Mellin space with $T_R=1/2$.
The important feature of Eq.\ (\ref{eq:9}) is that the hard
function $H_{ab}(M^2,\mu^2)$, contrary to the perturbative cross section
$\sigma_{ab}(N,M^2,b^2,\mu^2)$ in Eq.\ (\ref{eq:5}), no longer contains large
logarithms in either $\nbar$ or $\bbar$, since these have all been resummed in the
coefficient functions $C$, the evolution operators $E$, and the Sudakov exponent
$G$. The latter can be expanded as
\bea
 G_{ab}(M^2,\bar{N},\bar{b}^2,\mu^2)&=& L g_{ab}^{(1)}(\lambda)
 +g_{ab}^{(2)}(\lambda,\ln\bar{N},M^2/\mu^2)+~\cdots
 \label{eq:JR:G}
\eea
where $L=\ln\chi=\lambda/(a_S\beta_0)$ encodes the large logarithms in both
$\nbar$ and $\bbar$, $\beta_0=11C_A/6-2N_fT_R/3$ is the one-loop coefficient of
the QCD beta-function, and $N_f$ is the number of active quark flavors.
The first term in this expansion
\bea
 Lg_{ab}^{(1)}(\lambda)&=&{L\over 2\lambda\beta_0}
 (A_a^{(1)}+A_b^{(1)}) \big[2\lambda+\ln(1-2\lambda)\big]
 \label{eq:JR:g1}
\eea
collects the leading logarithmic (LL) contributions, while the second term
\bea
 2\beta_0g_{ab}^{(2)}(\lambda,\ln\bar{N},M^2/\mu^2)&=&
 (A_a^{(1)}+A_b^{(1)})\big[2\lambda\frac{1-2a_s\beta_0\ln\bar{N}}{1-2\lambda}
 +\ln(1-2\lambda)\big]\ln\frac{M^2}{\mu^2} \nonumber \\
 &+& (A_a^{(1)}+A_b^{(1)})\frac{\beta_1}{\beta_0^2}\big[
  (2\lambda+\ln(1-2\lambda))\frac{1-2a_s\beta_0\ln\bar{N}}{1-2\lambda}
  \!+\!\frac{1}{2}\ln^2(1-2\lambda)\big] \nonumber \\
 &-& (A_a^{(2)}+A_b^{(2)})
  \frac{1}{\beta_0}\big[2\lambda\frac{1-2a_s\beta_0\ln\bar{N}}{1-2\lambda}
  +\ln(1-2\lambda)\big] \nonumber\\
 &+& (-2\gamma_a^{(1)}-2\gamma_b^{(1)}+D_{ab}^{(1)})\ln(1-2\lambda)
 \label{eq:JR:g2}
\eea
with $\beta_1=(17C_A^2-5C_AN_f-3C_FN_f)/6$ collects the NLL contributions.
These terms reproduce those appearing in the transverse-momentum resummation
formalism \cite{Debove:2009ia}, when $\bbar\gg\nbar$, $\chi\to\bbar$ and
$(1-2a_s\beta_0\ln\nbar)/(1-2\lambda)\to 1/(1-2\lambda)$, as well as those
appearing in the threshold resummation formalism \cite{Debove:2010kf}, when
$\nbar\gg\bbar$, $\chi\to\nbar$ and $(1-2a_s\beta_0\ln\nbar)/(1-2\lambda)\to1$,
respectively. For transverse-momentum resummation, the logarithm ($\ln\bbar$)
had to be modified (to $\ln\sqrt{1+\bbar^2}$) in order to suppress unphysical
resummation contributions at large $p_T$ and small $\bbar$. This is not necessary
within the joint resummation formalism, as the small-$\bbar$ singularity is
regularized through the function $\chi(\nbar,\bbar)$. The coefficients
\bea
 A_a^{(1)}~=~ 2C_a &,& 
 A_a^{(2)}~=~ 2C_a\bigg[\bigg(\frac{67}{18}-\frac{\pi^2}{6}\bigg)C_A 
              -\frac{5}{9}N_f\bigg] ~~ {\rm and}~~
 D_{ab}^{(1)}~=~0
 \label{eq:TR:NllCoeff}
\eea
are well-known from both threshold and transverse-momentum resummation
\cite{Catani:1989ne}, while the anomalous dimensions
\bea
 \gamma_q^{(1)}~=~{3C_F\over2} &,& \gamma_g^{(1)}~=~\beta_0
\eea
of the quark and gluon fields have been introduced in order to remove the
corresponding NLL terms from the one-loop approximation of the diagonal evolution
operators $E_{aa}^{(1)}$.

While the large logarithms must clearly be resummed close to the production
threshold, when $z\to1$ and $\nbar\to\infty$, and/or at small values of $p_T\to0$,
when $\bbar\to\infty$, they account only
partially for the full perturbative cross section away from these regions. In
order to obtain a valid cross section at all values of $z$ and $p_T$, the
fixed-order (f.o.) and the resummed (res.) calculations must be combined
consistently by subtracting from their sum the perturbatively expanded (exp.)
resummed component,
\bea
 \sigma_{ab}&=&
 \sigma^{\rm(res.)}_{ab}+\sigma^{\rm(f.o.)}_{ab}-\sigma^{\rm(exp.)}_{ab}.
 \label{eq:JR:Mat}
\eea
The latter is easily obtained by expanding Eq.\ (\ref{eq:JR:pff}) to the desired
accuracy. At $\cO(a_s)$, one finds
\bea
 \sigma_{ab}^{\rm(exp)}(N,M^2,p_T^2,\mu^2)&=&
 H_{ab}^{(0)}(M^2,\mu^2) + a_s H_{ab}^{(1)}(M^2,\mu^2) \nonumber\\
&-& a_s\bigg(2\cJ -\ln \frac{M^2}{\mu^2}\bigg) \sum_c
 \big[H_{ac}^{(0)}(M^2,\mu^2) P_{cb}^{(1)}(N)
 \!+\!P_{ca}^{(1)}(N) H_{cb}^{(0)}(M^2,\mu^2)\big]\nonumber \\
&+& a_s \sum_c \big[H_{ac}^{(0)}(M^2,\mu^2) C_{cb}^{(1)}(N)
 +C_{ca}^{(1)}(N) H_{cb}^{(0)}(M^2,\mu^2)\big]\nonumber  \\
&-& a_s H_{ab}^{(0)}(M^2,\mu^2) \big[\cJ^2(A_a^{(1)}+A_b^{(1)})
  -2\cJ(\gamma_a^{(1)}+\gamma_b^{(1)})\big],
  \label{eq:JR:Tru}
\eea
where the $\cO(a_s,\eps^0)$ terms of the Altarelli-Parisi splitting functions are
\bea
 P_{qq}^{(1)}(N)&=&C_F\bigg[\frac{3}{2}+\frac{1}{N(N+1)}-2\sum_{k=1}^N\frac{1}{k}\bigg],
 \label{eq:pqqn}\\
 P_{gq}^{(1)}(N)&=&C_F\bigg[\frac{2+N+N^2}{N(N^2-1)}\bigg],\\
 P_{qg}^{(1)}(N)&=&T_R\bigg[\frac{2+N+N^2}{N(N+1)(N+2)}  \bigg],\\
 P_{gg}^{(1)}(N)&=&2C_A\bigg[\frac{1}{N(N-1)}+\frac{1}{(N+1)(N+2)}-\sum_{k=1}^N
             \frac{1}{k}\bigg]+\beta_0,
  \label{eq:pggn}
\eea
and the full dependence on the transverse momentum $p_T$ is embodied in the Bessel
integral
\bea
  \cJ&=&\int_0^\infty \d b\frac{b}{2}J_0(bp_T)\ln\chi(\bar{N},\bar{b}).
  \label{eq:JR:cI}
\eea
The form of this integral has been chosen to match the one encountered in
$p_T$-resummation. Unfortunately, it then induces $\pi^2$-terms which differ
slightly from those encountered in threshold resummation \cite{Laenen:2000ij}.
While these terms are formally beyond NLL order, one should nevertheless expect
slightly better numerical agreement with the $p_T$-resummed calculation than with
the threshold-resummed calculation.

After the resummation of the partonic cross section has been performed in $N$-
and $b$-space, we have to multiply the resummed cross section and its perturbative
expansion with the moments of the PDFs $f_{a/A}(N,\mu^2)$ and transform the
hadronic cross section obtained in this way back to the physical $z$- and
$p_T$-spaces. The moments of the PDFs are obtained through a numerical fit to the
publicly available PDF parameterizations in $x$-space. For the inverse
integral transforms, special attention has to be paid to the singularities in the
resummed exponents, {\em i.e.} when $\lambda=1/2$ in Eqs.\ (\ref{eq:JR:g1})
and (\ref{eq:JR:g2}). They are related to the presence of the Landau pole in the
perturbative running of $a_s(\mu^2)$, and prescriptions for both the Mellin and
Fourier inverse transforms are needed. For the Fourier inverse transform of Eq.\
(\ref{eq:JR:pff}), we follow Ref.\ \cite{Laenen:2000de} and deform the integration
contour of the $b$-integral in the complex plane by defining two integration
branches
\begin{equation}
  b=(\cos\varphi\pm i \sin\varphi)t,\quad t\in[0,\infty[,
  \label{eq:IT:bbra}
\end{equation}
where $\varphi$ has to be chosen in the range $]0,\pi/2[$.
The Bessel function $J_0(y)$ in Eq.\ (\ref{eq:JR:pff}) is then replaced by the sum
of the two auxiliary functions
\bea
 h_1(y,v)&=&-\frac{1}{2\pi}\int_{-iv\pi}^{-\pi+iv\pi}\d\theta e^{-iy\sin\theta},\\
 h_2(y,v)&=&-\frac{1}{2\pi}\int^{-iv\pi}_{\pi+iv\pi}\d\theta e^{-iy\sin\theta},
 \label{eq:IT:HanFun}
\eea
which are finite for any value of $y$. Their sum is independent of $v$ and is
always equal to $J_0(y)$. Since the two functions distinguish positive and
negative phases in the complex $b$-plane, they can be associated with only one of
the two branches. For the inverse Mellin transform
\bea
  F(y) &=& \int_{\cC_N} \frac{\d N}{2\pi i} y^{-N} F(N)
  \label{eq:IT:InvMel}
\eea
we choose an integration contour $\cC_N$ according to the principal value
procedure proposed in Ref.\ \cite{Contopanagos:1993yq} and the minimal
prescription proposed in Ref.\ \cite{Catani:1996yz} and define again two branches
\bea
  \cC_N:\quad N&=&C+ye^{\pm i\phi},\quad y\in[0,\infty[.
  \label{eq:IT:Nbra}
\eea
The parameter $C$ must be chosen in such a way that the poles in the Mellin
moments of the parton densities, which are related to the small-$x$ (Regge)
singularity $f_{a/A}(x,\mu_0^2)\propto x^\alpha(1-x)^\beta$ with $\alpha<0$,
lie to the left and the Landau pole to the right of the integration contour,
respectively. While formally the angle $\phi$ can be chosen in the range
$[\pi/2,\pi[$, it is advantageous to take $\phi>\pi/2$ to improve the convergence
of the inverse Mellin transform.

\section{Numerical results}
\label{sec:3}

We now turn to our numerical analysis of joint resummation effects on the
production of various gaugino pairs at the Tevatron $p\bar{p}$-collider
($\sqrt{S}=1.96$ TeV) and the LHC $pp$-collider ($\sqrt{S}=7-14$ TeV). For the
masses and widths of the electroweak gauge bosons, we use the values of
$m_Z=91.1876$ GeV and $m_W=80.403$ GeV
\cite{Amsler:2008zzb}. The CKM-matrix is assumed to be diagonal, and
the top quark mass is taken to be 173.1 GeV \cite{:2009ec}. The strong coupling
constant is evaluated in the one-loop and two-loop approximation for LO and
NLO/NLL+NLO results, respectively, with a value of $\Lambda_{\overline{\rm MS}}^
{n_f=5}$ corresponding to the employed LO (CTEQ6L1) and NLO (CTEQ6.6M) parton
densities \cite{Nadolsky:2008zw}. For the resummed and expanded contributions, the
latter have been transformed numerically to Mellin $N$-space. When we
present spectra in the invariant mass $M$ of the gaugino pair, we identify
the unphysical scales $\mu_{F}=\mu_R=\mu$ with $M$, whereas for transverse
momentum distributions and total cross sections we identify them with the average
mass of the two produced gauginos. The remaining theoretical uncertainty is
estimated by varying the common scale $\mu$ about these central values by a factor
of two up and down. The running electroweak couplings as well as the
physical masses of the SUSY particles and their mixing angles are computed
with the computer program SPheno 2.2.3 \cite{Porod:2003um}, which includes a
consistent calculation of the Higgs boson masses and all one-loop and
the dominant two-loop radiative corrections in the renormalization group
equations linking the restricted set of SUSY-breaking parameters at the gauge
coupling unification scale to the complete set of observable SUSY masses and
mixing angles at the electroweak scale.

For the Tevatron, we choose the low-mass point LM0 (SU4) with universal fermion
mass $m_{1/2}=160$ GeV, scalar mass $m_0=200$ GeV, trilinear coupling $A_0=-400$
GeV, bilinear Higgs mass parameter $\mu>0$, and ratio of Higgs vacuum expectation
values $\tan\beta=10$ \cite{Khachatryan:2011tk,Aad:2009wy}. It has been been
defined by the CMS
(ATLAS) collaboration with the objective of high cross sections and thus early
discovery at the LHC, as the resulting gaugino, squark and slepton masses $\mn=
\mc=113$ GeV, $\md=61$ GeV, $\ms\simeq420$ GeV, and $\ml\simeq220$ GeV lie just
beyond the current Tevatron limits. In this scenario, the lightest chargino and
second-lightest neutralino decay with 35\% and 15\% probability through virtual
sleptons to the LSP and one and two charged leptons, respectively
\cite{Muhlleitner:2003vg}.

For the LHC, we choose the widely used minimal supergravity (mSUGRA) point SPS1a'
\cite{AguilarSaavedra:2005pw} as the benchmark for our numerical studies. This
point has the same intermediate value of $\tan\beta=10$ and $\mu>0$ (favored by
the rare decay $b\to s\gamma$ and the measured anomalous magnetic moment of the
muon), a still relatively small gaugino mass parameter of $m_{1/2}
=250$ GeV, and a slightly lower scalar mass parameter $m_0=70$ GeV and trilinear
coupling $A_0=-300$ GeV than the original point SPS1a \cite{Allanach:2002nj} in
order to render it compatible with low-energy precision data, high-energy mass
bounds, and the observed cold dark matter relic density. It is also similar to
the post-WMAP point B' ($m_0=60$ GeV and $A_0=0$) \cite{Battaglia:2003ab}, which
has been adopted by the CMS collaboration as their low-mass point LM1
\cite{Ball:2007zza}. In the SPS1a' scenario, the $\na$ is the LSP with a mass of
98 GeV, the gauginos producing the trilepton signal have masses of $m_{\ca}\simeq
m_{\nb}=184$ GeV, and the heavier gauginos, which decay mostly into the lighter
gauginos, $W$ and $Z$ bosons as well as the lightest Higgs boson, have masses of
$m_{\nc}=400$ GeV and
$m_{\cb}\simeq m_{\nd}=415$ GeV. The average squark and gluino masses are
$m_{\tilde{q}}\simeq 550$ GeV and $\mgl=604$ GeV. Note that this benchmark point
is also relatively close to the region excluded by the Tevatron
collaborations CDF and D0, which assume, however, a lower value of $\tan\beta=3$
and $A_0=0$ \cite{Aaltonen:2008pv}.

\subsection{Transverse momentum spectra}

\begin{figure}
 \centering
 \epsfig{file=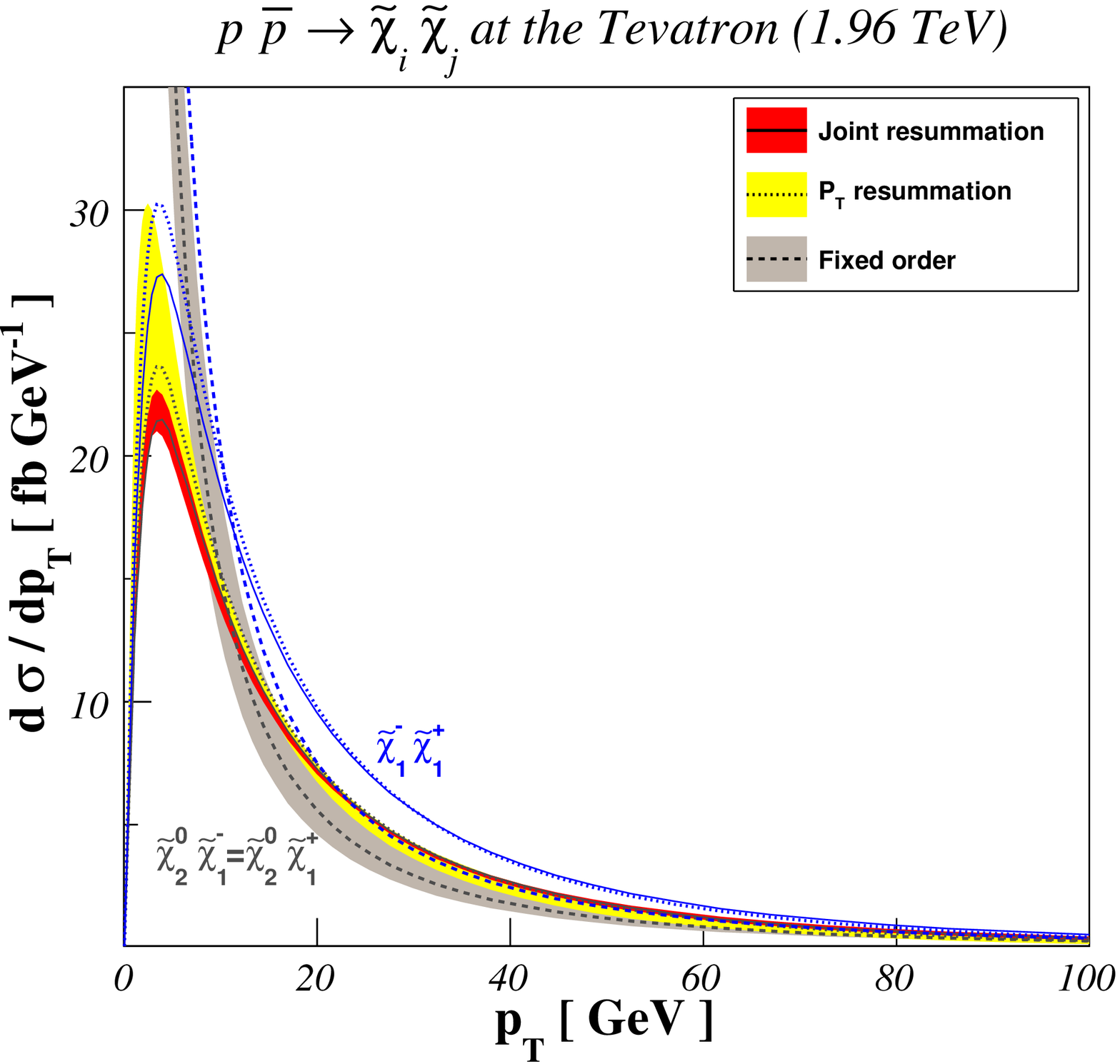,width=.65\columnwidth}
 \epsfig{file=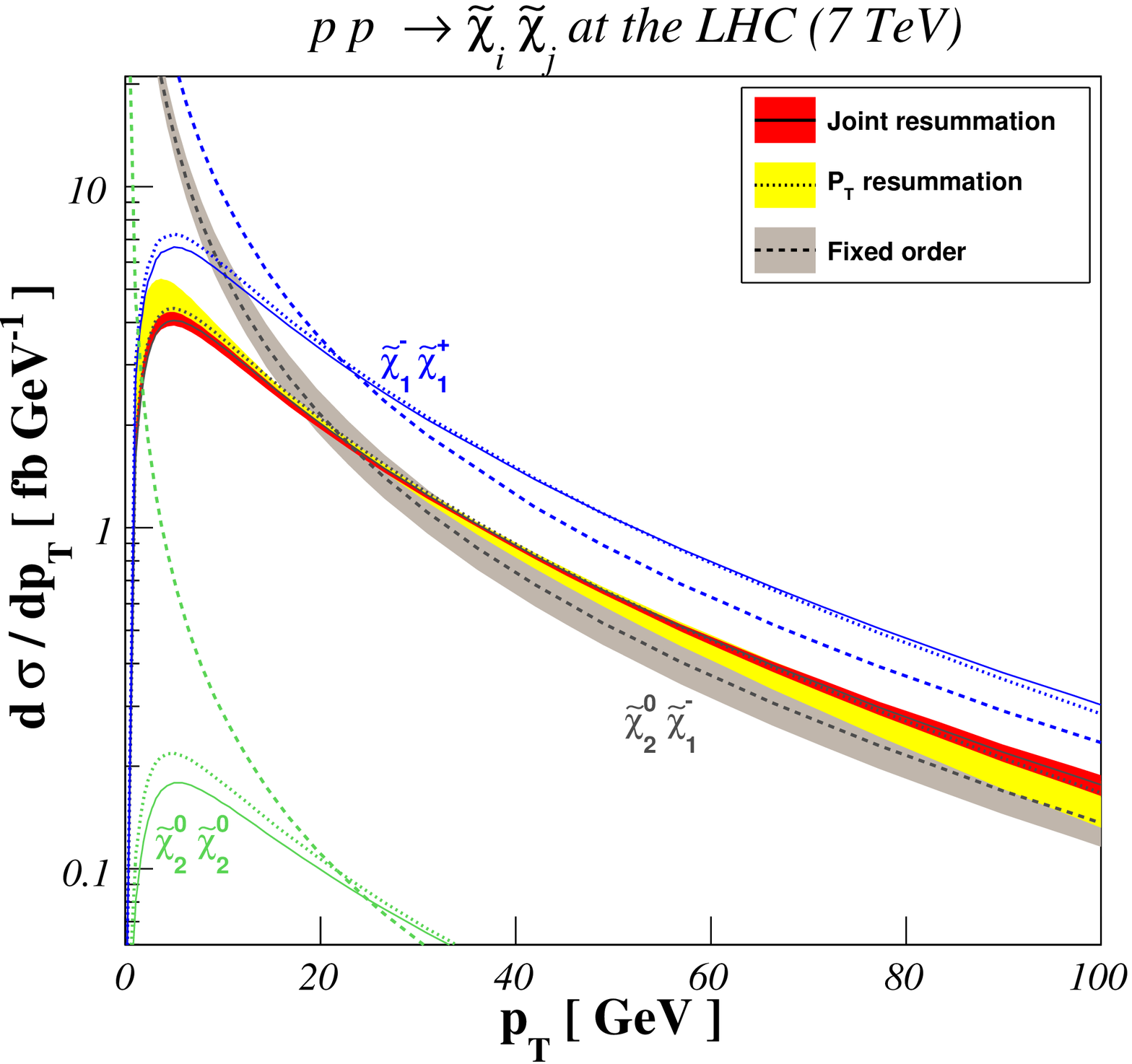,width=.65\columnwidth}
 \caption{\label{fig:1}Transverse momentum distributions of light gaugino pairs
 at the Tevatron (top) and LHC (bottom) with $\sqrt{S}=1.96$ and 7 TeV
 center-of-mass energy, respectively, in fixed order (dashed) as well as with
 transverse-momentum (dotted) and joint (full curves) resummation.}
\end{figure}

In Fig.\ \ref{fig:1}, we present transverse momentum spectra for light charged and
%
%
neutral gaugino pairs with masses of 113 and 184 GeV at the Tevatron (top)
and early LHC (bottom), where the center-of-mass energies are $\sqrt{S}=1.96$ and
7 TeV, respectively. We show predictions at fixed order ${\cal O}(\alpha_s)$
(dashed) as well as with transverse-momentum (dotted) and joint resummation
(full curves). While the fixed-order predictions diverge at small $p_T$ due to an
uncancelled soft singularity from real gluon emission, the resummed predictions
exhibit a finite, physical behavior with a pronounced maximum in the region of
$p_T=5$ to 10 GeV. In the region of intermediate $p_T$ of 20 to 60 GeV, the
resummed predictions are considerably larger than those at fixed order. The
calculations using transverse-momentum and joint resummation are in good
agreement, but the theoretical uncertainty of the latter, estimated by varying
the renormalization and factorization scales by a factor of two about the
average mass of the two gauginos, is considerably smaller, since threshold
logarithms are resummed simultaneously. The cross sections for chargino pairs
exceed those for the trilepton channel at the Tevatron and for the $\nb\cc$
channel (but not the $\nb\cf$ channel, which is not shown) at the early LHC with
the cross section for pair production of the second-lightest neutralino being more
than an order of magnitude smaller at the early LHC.

\begin{figure}
 \centering
 \epsfig{file=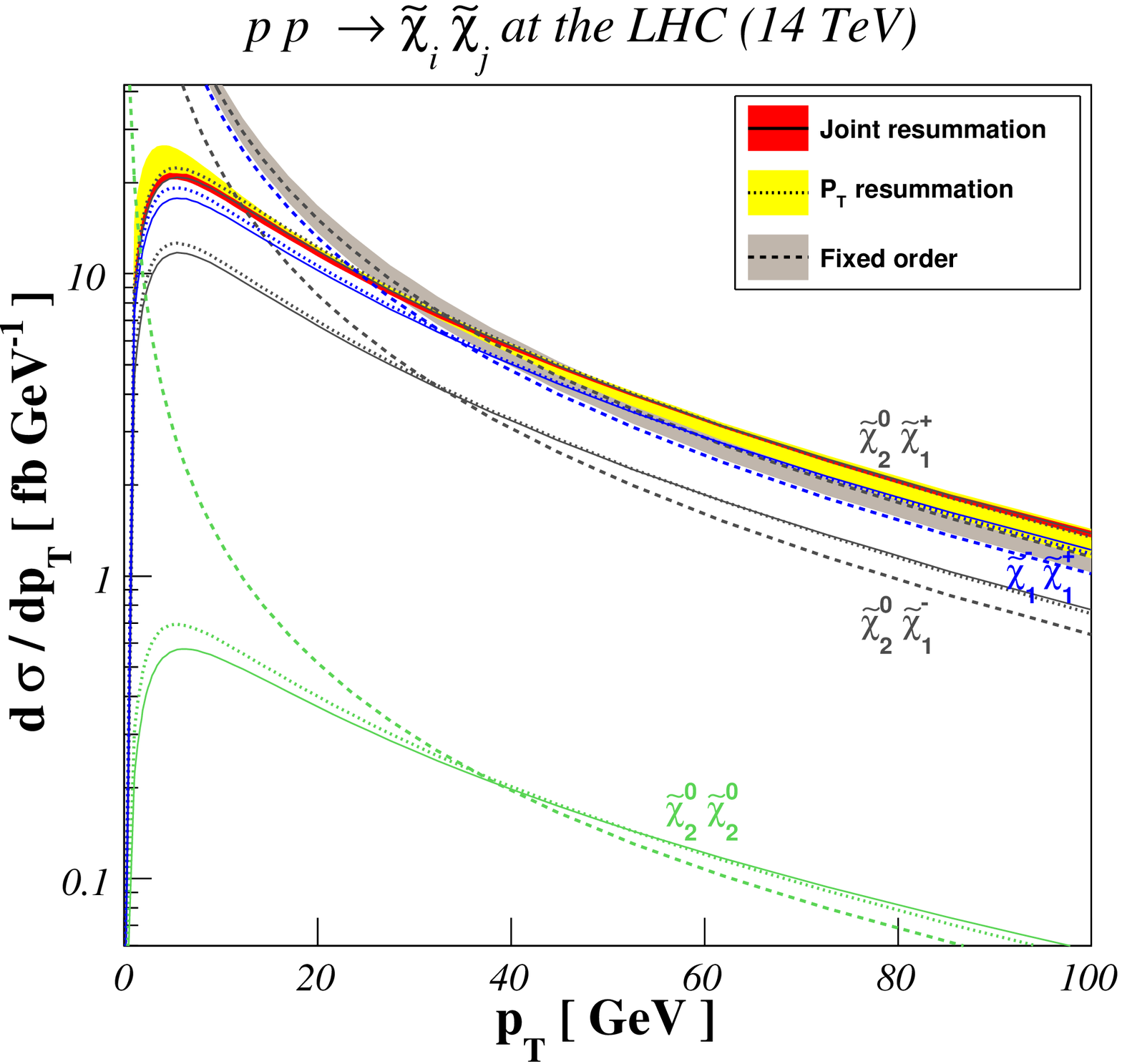,width=.65\columnwidth}
 \epsfig{file=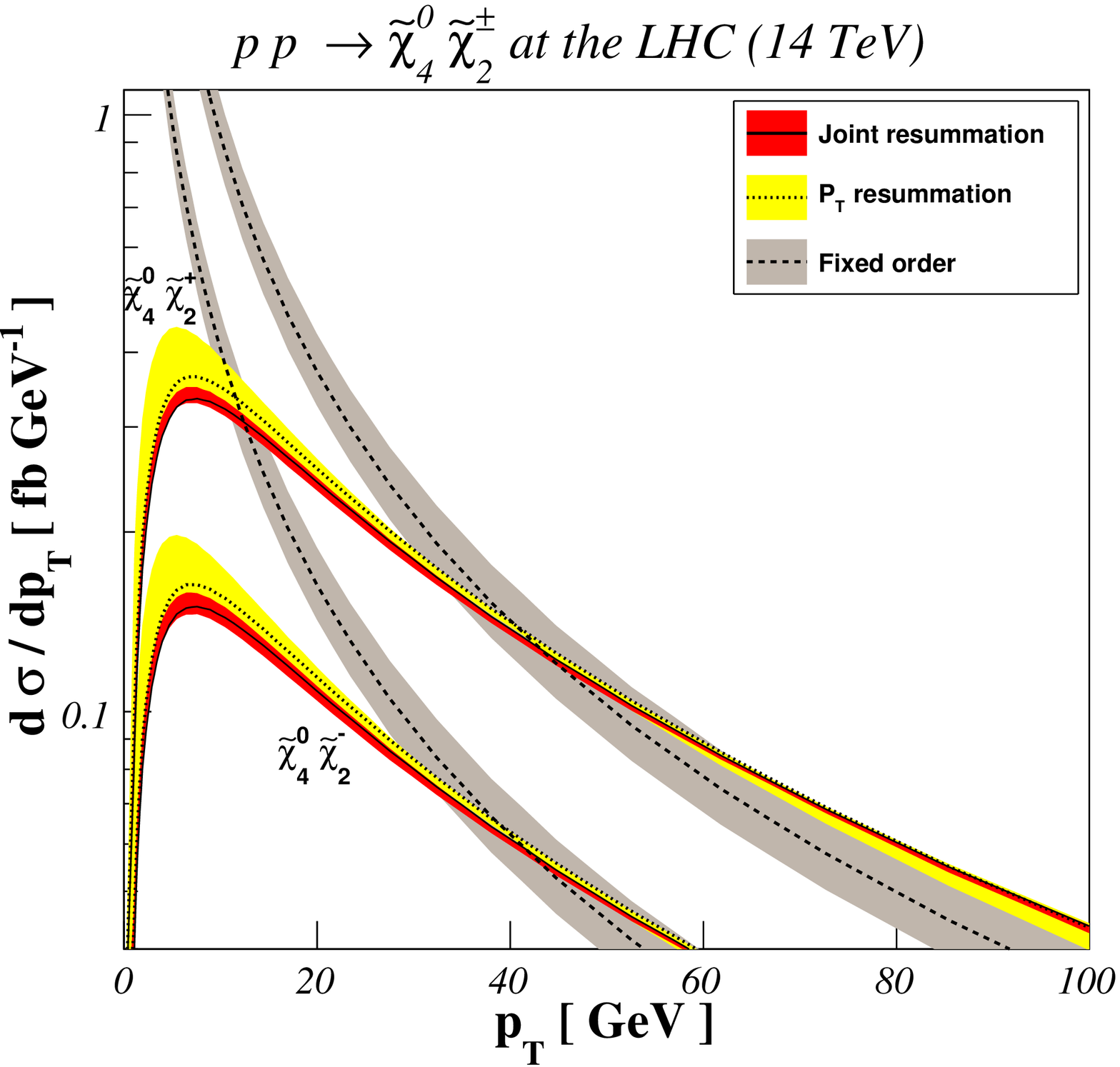,width=.65\columnwidth}
 \caption{\label{fig:2}Transverse momentum distributions of light (top) and
 heavy (bottom) gaugino pairs at the LHC with $\sqrt{S}=14$ TeV center-of-mass
 energy in fixed order (dashed) as well as with transverse-momentum (dotted) and
 joint (full curves) resummation.}
\end{figure}

At the LHC design luminosity of $\sqrt{S}=14$ TeV, the neutralino pair production
cross section is larger by more than a factor of three (see the top part of Fig.\
\ref{fig:2}). Otherwise, the behavior of the various predictions is very similar
%
%
to the one described above. One notices, however, that the resummed predictions
exceed those at fixed order only at larger values of $p_T>40$ GeV. Furthermore,
the trilepton cross section for positive charginos is larger than the one for
negative charginos by almost a factor of two, since in contrast to the Tevatron
the LHC is a $pp$ collider. With a center-of-mass energy of $\sqrt{S}=14$ TeV, it
may also become possible to observe the associated production of heavier
neutralinos $\nd$ and charginos $\cb$ with masses of about 415 GeV. The
corresponding transverse momentum spectra are shown in the lower part of Fig.\
\ref{fig:2}. The absolute values of the cross sections are reduced by about a
factor of 50 for both positive and negative charginos, but the shape of the
distributions is very similar. Joint resummation leads again to the smallest
scale uncertainties.

\subsection{Invariant mass spectra}

While the joint resummation formalism is designed to match more closely the one
for transverse-momentum resummation, it also allows to simultaneously resum
threshold logarithms and obtain precise invariant mass spectra. These are
therefore presented in this section for various gaugino pairs and colliders and
compared to those obtained with pure threshold resummation with the expectation
that the agreement will be slightly worse than the one for transverse momentum
distributions.

\begin{figure}
 \centering
 \epsfig{file=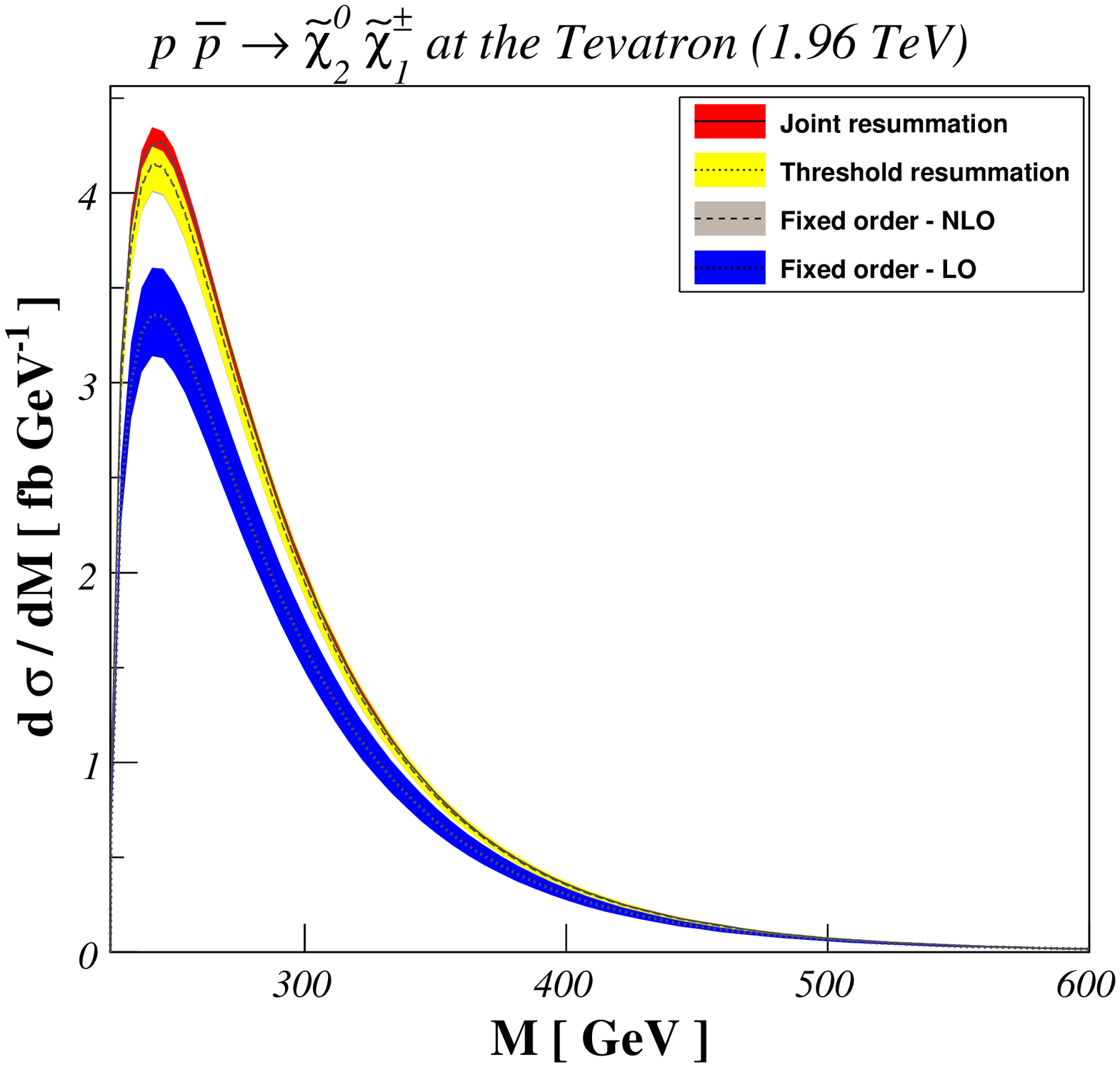,width=.65\columnwidth}
 \caption{\label{fig:3}Invariant mass distributions of light gaugino pairs
 at the Tevatron with $\sqrt{S}=1.96$ TeV center-of-mass energy in fixed order
 (short-dashed, dashed) as well as with threshold (dotted) and joint (full curve)
 resummation.}
\end{figure}

In Fig.\ \ref{fig:3} we show invariant mass spectra for the trilepton channel
%
%
at the Tevatron obtained at LO (short-dashed) and with NLO SUSY-QCD corrections
(dashed) as well as with pure threshold (dotted) and joint (full curve)
resummation. In this figure, the NLO and threshold resummed predictions are both
considerably larger than the one obtained at LO. In the linear representation
of $\d\sigma/\d M$ emphasizing the low-invariant mass region shown here, they
can in fact not be distinguished, as threshold effects only start to dominate as
the invariant mass squared $M^2$ approaches the total available center-of-mass
energy $\sqrt{s}$. This is also the reason why the jointly resummed prediction
differs and in fact exceeds slightly the pure threshold resummed prediction at
small $M$, as large logarithms at small $p_T$ have been simultaneously resummed.
This leads also to an additional reduction of the scale uncertainty, represented
again as a shaded band and obtained by varying the renormalization and
factorization scales by a factor of two about the invariant mass $M$.

The various features described above are even more prominent at the LHC with its
larger design center-of-mass energy of $\sqrt{S}=14$ TeV, despite the fact that
also the masses of the light gauginos are slightly larger in the SPS1a' scenario
than at the LM0 benchmark point. The scale uncertainties in the upper part of
%
\begin{figure}
 \centering
 \epsfig{file=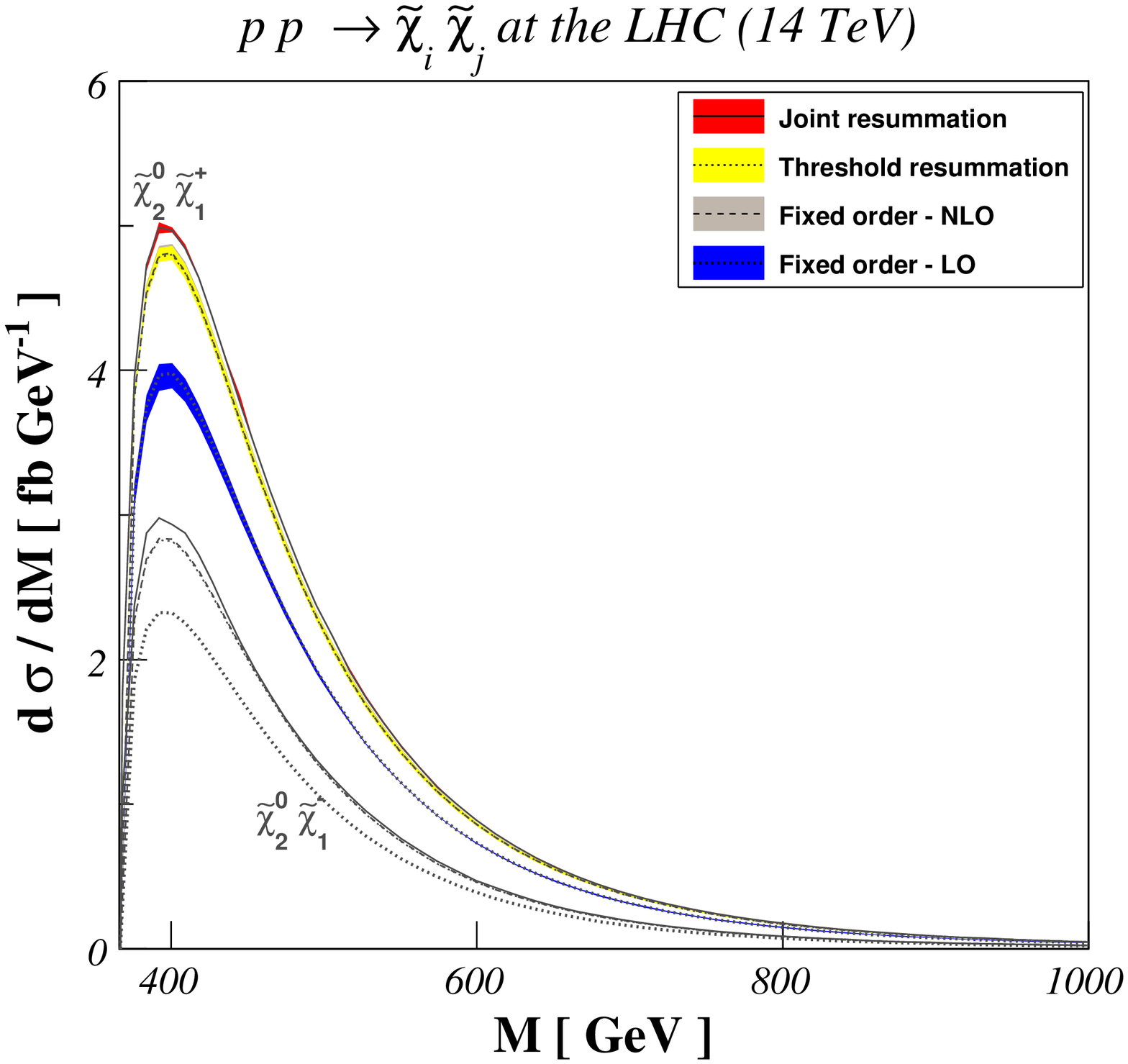,width=.65\columnwidth}
 \epsfig{file=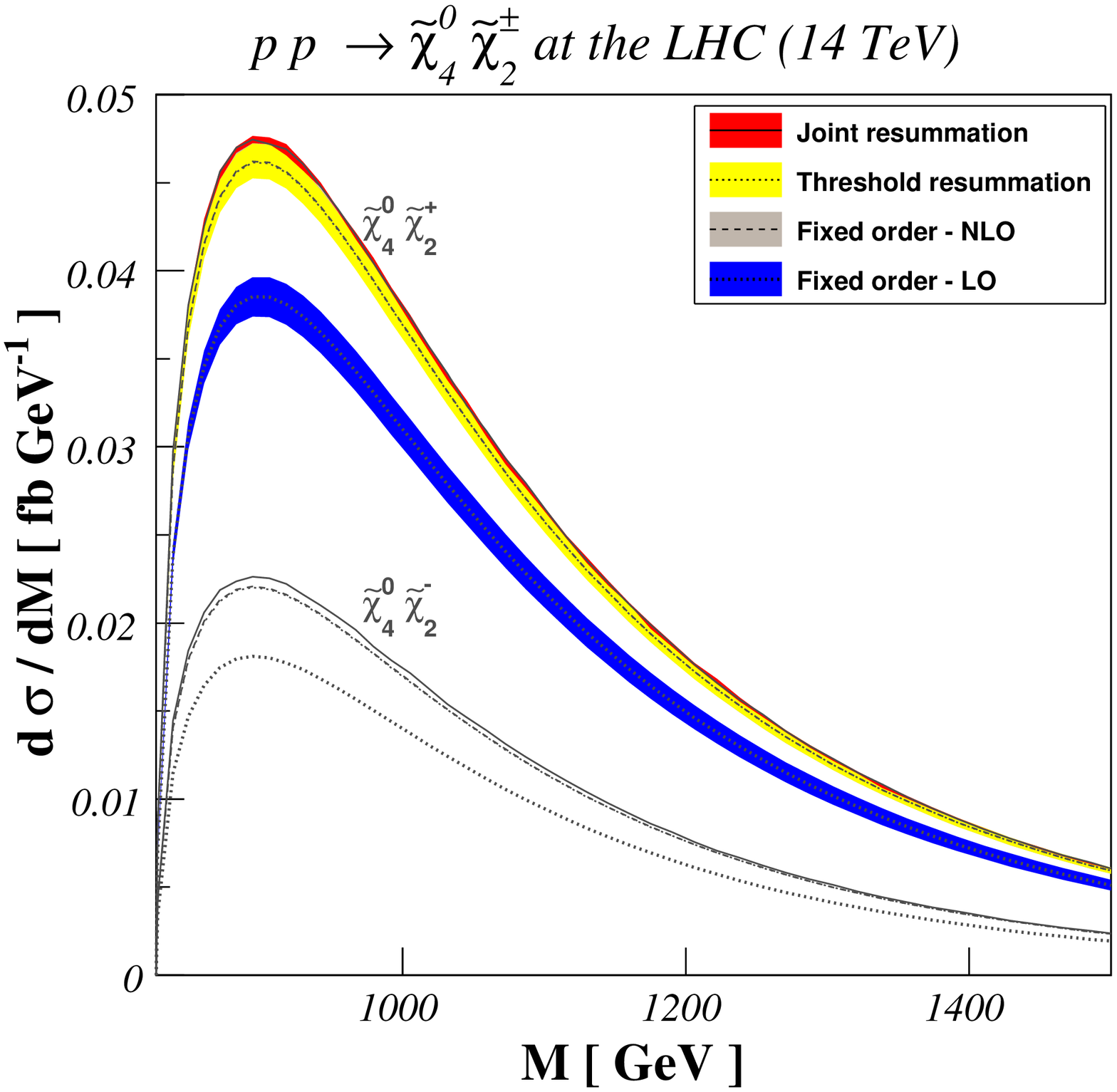,width=.65\columnwidth}
 \caption{\label{fig:4}Invariant mass distributions of light (top) and heavy
 (bottom) gaugino pairs at the LHC with $\sqrt{S}=14$ TeV center-of-mass energy
 in fixed order (short-dashed, dashed) as well as with threshold (dotted)
 and joint (full) resummation.}
\end{figure}
%
Fig.\ \ref{fig:4} are considerably smaller at LO and even more at NLO and NLL+NLO,
as the variation from $M/2$ to $2M$ is less important compared to the large
$\sqrt{S}$. This leads to threshold and joint resummation predictions that no
longer overlap within scale uncertainties in the region of maximal cross section,
indicating that the theoretical error is slightly underestimated in this case. As
it was already observed in the previous section, the cross section for the
trilepton channel with positive charge exceeds the negative one by almost a factor
of two due to the positively charged initial state at the LHC. Heavier
gauginos are produced closer to threshold, so that the distributions in the
lower part of Fig.\ \ref{fig:4} are found at larger values of $M$. The scale
variation and the shaded bands representing it become more important again with
respect to $\sqrt{S}$ as do the threshold logarithms, so that the threshold
and jointly resummed predictions overlap again within the theoretical
uncertainties. The absolute size of the cross section is again almost two times
larger for $\nd\ce$ pairs than for $\nd\cd$ pairs.

\subsection{Total cross sections}

Total cross sections can be obtained from the distributions shown in the
previous sections by integrating over either $p_T$ or $M$. This is numerically
not always trivial, but should in principle lead to similar predictions, building
confidence in the theoretical calculations. In particular, since the perturbative,
resummed and expanded contributions are obtained in $(p_T,~M)$ and $(b,~N)$ space,
respectively, we first compute the differential cross section $\d\sigma/
\d p_T$ and then integrate over $p_T$ with the trapezoidal rule only after
the different contributions to the total cross section have been matched.

The results are displayed in Tab.\ \ref{tab:1}. They show the expected important
increase in absolute size and
\begin{table}
\caption{\label{tab:1}Total cross sections (in fb) with scale uncertainties for
 the trilepton channel at the Tevatron in the LM0 scenario and for light and heavy
 gaugino pairs at the LHC with 7 and 14 TeV center-of-mass energy in the SPS 1a'
 scenario at various levels of accuracy.\\}
\begin{tabular}{|c|c|ccccc|}
\hline
Collider & Gauginos & LO & NLO & $p_T$ & Threshold & Joint \\
\hline
Tevatron     & $\nb\ca$ &
        $300.044^{+27.555}_{-23.707}$ &
        $366.794^{+10.772}_{-11.750}$ &
        $377.118^{+ 5.891}_{-14.548}$ &
        $363.922^{+ 1.702}_{- 2.801}$ &
        $365.974^{+ 3.542}_{- 3.847}$ \\
LHC-7~~                & $\nb\cc$ &
        $102.245^{+3.494}_{-3.564}$ &
        $121.216^{+1.843}_{-1.557}$ &
        $122.817^{+1.338}_{-1.302}$ &
        $119.885^{+0.118}_{-0.632}$ &
        $121.188^{+0.084}_{-0.590}$ \\
LHC-14              & $\nb\cc$ &
        $346.538^{+2.210}_{-4.862}$ &
        $419.930^{+4.016}_{-1.809}$ &
        $429.678^{+0.962}_{-1.605}$ &
        $416.327^{+0.895}_{-2.547}$ &
        $428.202^{+0.464}_{-2.680}$ \\
LHC-14            & $\nd\cd$ &
        $6.506^{+0.308}_{-0.291}$ &
        $7.844^{+0.140}_{-0.133}$ &
        $8.261^{+0.179}_{-0.174}$ &
        $7.763^{+0.017}_{-0.008}$ &
        $8.117^{+0.033}_{-0.047}$\\
\hline
\end{tabular}
\end{table}
reduction in scale uncertainty from LO to NLO and then, to a lesser extent,
at NLL+NLO. It is interesting to note that the $p_T$-resummed predictions
are indeed larger than those obtained at NLO, but the threshold resummed
predictions are slightly smaller. The jointly resummed predictions are very
similar to those obtained at NLO at the two lower center-of-mass energies and larger at the LHC with
$\sqrt{S}=14$ TeV. As expected, the lowest scale uncertainties are found with
threshold resummation when the particle masses are large compared to the
available center-of-mass energy (light gauginos at the Tevatron,
heavy gauginos at the LHC with $\sqrt{S}=$ 14 TeV), while joint resummation
gives results that are similar to $p_T$ resummation, but more precise, in the
two other cases.

\section{Conclusion}
\label{sec:4}

In this paper, we have completed our investigation of gaugino production at
hadron colliders with different resummation methods by presenting a NLL+NLO
calculation that jointly resums large logarithms in the small-$p_T$ and
threshold regions. After a detailed outline of the organization of the
analytical calculation and the numerical implementation, in particular of
the convolutions with the parton densities and the inverse Fourier and
Mellin transforms, we have compared the new jointly resummed predictions to
those obtained previously at fixed order as well as with pure $p_T$ and
threshold resummation. We found in general good agreement in the transverse
momentum and invariant mass distributions, confirming the reliability of the
joint resummation method. In most cases the new results were also even more
precise, so that they can be considered the most reliable predictions for
direct gaugino production available for the Tevatron and the LHC. They are
therefore of great importance for supersymmetry searches and parameter
determinations at these hadron colliders and should be taken as the basis
for future experimental analyses.

\acknowledgments

We thank E.\ Conte and Y.\ Patois for their help with using the grid.
This work has been supported by a Ph.D.\ fellowship of the French ministry
for education and research and by the Theory-LHC-France initiative of the
CNRS/IN2P3.



\begin{thebibliography}{00}


\bibitem{Nilles:1983ge}
  H.~P.~Nilles,
  Phys.\ Rept.\  {\bf 110} (1984) 1;
%
  H.~E.~Haber and G.~L.~Kane,
  Phys.\ Rept.\  {\bf 117} (1985) 75;
%
  J.~F.~Gunion and H.~E.~Haber,
  Nucl.\ Phys.\  B {\bf 272} (1986) 1
  [Erratum-ibid.\  B {\bf 402} (1993) 567].


\bibitem{Aaltonen:2008pv}
  T.~Aaltonen {\it et al.}  [CDF Collaboration],
  Phys.\ Rev.\ Lett.\  {\bf 101} (2008) 251801;
%
  R.~Forrest  [CDF Collaboration],
  arXiv:0910.1931 [hep-ex];
%
  V.~M.~Abazov {\it et al.}  [D0 Collaboration],
  Phys.\ Lett.\  B {\bf 680} (2009) 34.


\bibitem{Aad:2009wy}
  G.~Aad {\it et al.}  [ATLAS Collaboration],
  arXiv:0901.0512.

\bibitem{Ball:2007zza}
  G.~Bayatian {\it et al.}  [CMS Collaboration],
  J.\ Phys.\ G {\bf 34} (2007) 995.


\bibitem{Barger:1983wc}
  V.~Barger, R.~Robinett, W.~Keung and R.~Phillips,
  Phys.\ Lett.\  B {\bf 131} (1983) 372;
%
  S.~Dawson, E.~Eichten and C.~Quigg,
  Phys.\ Rev.\  D {\bf 31} (1985) 1581;
%
  D.~A.~Dicus, S.~Nandi and J.~Woodside,
  Phys.\ Rev.\  D {\bf 41} (1990) 2347;
%
%
  M.~Klasen and G.~Pignol,
  Phys.\ Rev.\  D {\bf 75} (2007) 115003.


\bibitem{Bozzi:2005sy}
  G.~Bozzi, B.~Fuks and M.~Klasen,
  Phys.\ Rev.\  D {\bf 72} (2005) 035016;
%
  D.~Berdine and D.~Rainwater,
  Phys.\ Rev.\  D {\bf 72} (2005) 075003;
%
  S.~Bornhauser, M.~Drees, H.~K.~Dreiner and J.~S.~Kim,
  Phys.\ Rev.\  D {\bf 76} (2007) 095020.


\bibitem{Gehrmann:2004xu}
  T.~Gehrmann, D.~Maitre and D.~Wyler,
  Nucl.\ Phys.\  B {\bf 703} (2004) 147;
%
  G.~Bozzi, B.~Fuks and M.~Klasen,
  Phys.\ Lett.\  B {\bf 609} (2005) 339;
%
  J.~Debove, B.~Fuks and M.~Klasen,
  Phys.\ Rev.\  D {\bf 78} (2008) 074020;
%
  M.~Klasen,
  arXiv:1005.3503.


\bibitem{Bozzi:2007me}
  G.~Bozzi, B.~Fuks, B.~Herrmann and M.~Klasen,
  Nucl.\ Phys.\  B {\bf 787} (2007) 1;
%
  F.~del Aguila {\it et al.},
  Eur.\ Phys.\ J.\  C {\bf 57} (2008) 183;
%
  B.~Fuks, B.~Herrmann and M.~Klasen,
  Nucl.\ Phys.\  B {\bf 810} (2009) 266.

\bibitem{Alan:2007rp}
  A.~T.~Alan, K.~Canko\c{c}ak and D.~A.~Demir,
  Phys.\ Rev.\  D {\bf 75} (2007) 095002
  [Erratum-ibid.\  D {\bf 76} (2007) 119903].


\bibitem{Beenakker:1996ch}
  W.~Beenakker, R.~H\"opker, M.~Spira and P.~M.~Zerwas,
  Nucl.\ Phys.\  B {\bf 492} (1997) 51;
%
  W.~Beenakker, M.~Kr\"amer, T.~Plehn, M.~Spira and P.~M.~Zerwas,
  Nucl.\ Phys.\  B {\bf 515} (1998) 3;
%
  H.~Baer, B.~W.~Harris and M.~H.~Reno,
  Phys.\ Rev.\  D {\bf 57} (1998) 5871;
%
  E.~L.~Berger, M.~Klasen and T.~M.~P.~Tait,
  Phys.\ Rev.\  D {\bf 59} (1999) 074024;
%
%
  W.~Beenakker, M.~Klasen, M.~Kr\"amer, T.~Plehn, M.~Spira and P.~M.~Zerwas,
  Phys.\ Rev.\ Lett.\  {\bf 83} (1999) 3780
  [Erratum-ibid.\  {\bf 100} (2008) 029901];
%
  E.~L.~Berger, M.~Klasen and T.~M.~P.~Tait,
  Phys.\ Lett.\  B {\bf 459} (1999) 165;
%
  E.~L.~Berger, M.~Klasen and T.~M.~P.~Tait,
  Phys.\ Rev.\  D {\bf 62} (2000) 095014
  [Erratum-ibid.\  {\bf 67} (2003) 099901];
%
  M.~Spira,
  arXiv:hep-ph/0211145;
%
  L.~G.~Jin, C.~S.~Li and J.~J.~Liu,
  Eur.\ Phys.\ J.\  C {\bf 30} (2003) 77;
%
  L.~G.~Jin, C.~S.~Li and J.~J.~Liu,
  Phys.\ Lett.\  B {\bf 561} (2003) 135.


\bibitem{Hollik:2007wf}
  W.~Hollik, M.~Kollar and M.~K.~Trenkel,
  JHEP {\bf 0802} (2008) 018;
%
  W.~Hollik, E.~Mirabella and M.~K.~Trenkel,
  JHEP {\bf 0902} (2009) 002;
%
  E.~Mirabella,
  JHEP {\bf 0912} (2009) 012.


\bibitem{Bozzi:2006fw}
  G.~Bozzi, B.~Fuks and M.~Klasen,
  Phys.\ Rev.\  D {\bf 74} (2006) 015001;
%
  M.~Klasen,
  Nucl.\ Phys.\ Proc.\ Suppl.\  {\bf 160} (2006) 111;
%
  L.~L.~Yang, C.~S.~Li, J.~J.~Liu and Q.~Li,
  Phys.\ Rev.\  D {\bf 72} (2005) 074026;
%
  H.~K.~Dreiner, S.~Grab, M.~Kr\"amer and M.~K.~Trenkel,
  Phys.\ Rev.\  D {\bf 75} (2007) 035003;
%
  Y.~Q.~Chen, T.~Han and Z.~G.~Si,
  JHEP {\bf 0705} (2007) 068.

\bibitem{Debove:2009ia}
  J.~Debove, B.~Fuks and M.~Klasen,
  Phys.\ Lett.\  B {\bf 688} (2010) 208;
%
  J.~Debove,
  proceedings of the 2009 Europhysics Conference on {\em High Energy Physics} (EPS
  HEP 2009), Cracow, Poland, arXiv:0908.4149 [hep-ph].


\bibitem{Bozzi:2007qr}
  G.~Bozzi, B.~Fuks and M.~Klasen,
  Nucl.\ Phys.\  B {\bf 777} (2007) 157;
%
  C.~S.~Li, Z.~Li, R.~J.~Oakes and L.~L.~Yang,
  Phys.\ Rev.\  D {\bf 77} (2008) 034010;
%
  A.~Kulesza and L.~Motyka,
  Phys.\ Rev.\ Lett.\  {\bf 102} (2009) 111802;
%
  U.~Langenfeld and S.~O.~Moch,
  Phys.\ Lett.\  B {\bf 675} (2009) 210;
%
  A.~Kulesza and L.~Motyka,
  Phys.\ Rev.\  D {\bf 80} (2009) 095004;
%
  W.~Beenakker, S.~Brensing, M.~Kr\"amer, A.~Kulesza, E.~Laenen and I.~Niessen,
  JHEP {\bf 0912} (2009) 041;
%
  W.~Beenakker, S.~Brensing, M.~Kr\"amer, A.~Kulesza, E.~Laenen and I.~Niessen,
  JHEP {\bf 1008} (2010) 098.

\bibitem{Debove:2010kf}
  J.~Debove, B.~Fuks and M.~Klasen,
  Nucl.\ Phys.\  B {\bf 842} (2011) 51;
%
  J.~Debove,
  proceedings of the 2010 Rencontres de Moriond on {\em QCD and High-Energy
  Interactions} (Moriond QCD 2010), La Thuile, Italy, arXiv:1009.2436 [hep-ph].


\bibitem{Li:1998is}
  H.~n.~Li,
  Phys.\ Lett.\  B {\bf 454} (1999) 328.

\bibitem{Laenen:2000de}
  E.~Laenen, G.~F.~Sterman and W.~Vogelsang,
  Phys.\ Rev.\ Lett.\  {\bf 84} (2000) 4296.

\bibitem{Laenen:2000ij}
  E.~Laenen, G.~F.~Sterman and W.~Vogelsang,
  Phys.\ Rev.\  D {\bf 63} (2001) 114018;
%
  A.~Kulesza, G.~F.~Sterman and W.~Vogelsang,
  Phys.\ Rev.\  D {\bf 66} (2002) 014011;
%
  A.~Kulesza, G.~F.~Sterman and W.~Vogelsang,
  Phys.\ Rev.\  D {\bf 69} (2004) 014012;
%
  A.~Banfi and E.~Laenen,
  Phys.\ Rev.\  D {\bf 71} (2005) 034003;
%
  G.~Bozzi, B.~Fuks and M.~Klasen,
  Nucl.\ Phys.\  B {\bf 794} (2008) 46;
%
  B.~Fuks, M.~Klasen, F.~Ledroit, Q.~Li and J.~Morel,
  Nucl.\ Phys.\  B {\bf 797} (2008) 322.


\bibitem{Altarelli:1977zs}
  G.~Altarelli and G.~Parisi,
  Nucl.\ Phys.\  B {\bf 126} (1977) 298.

\bibitem{Furmanski:1981cw}
  W.~Furmanski and R.~Petronzio,
  Z.\ Phys.\  C {\bf 11} (1982) 293.


\bibitem{Catani:1989ne}
  S.~Catani and L.~Trentadue,
  Nucl.\ Phys.\  B {\bf 327} (1989) 323;
%
  S.~Catani and L.~Trentadue,
  Nucl.\ Phys.\  B {\bf 353} (1991) 183.


\bibitem{Contopanagos:1993yq}
  H.~Contopanagos and G.~Sterman,
  Nucl.\ Phys.\  B {\bf 419} (1994) 77.

\bibitem{Catani:1996yz}
  S.~Catani, M.~L.~Mangano, P.~Nason and L.~Trentadue,
  Nucl.\ Phys.\  B {\bf 478} (1996) 273.


\bibitem{Amsler:2008zzb}
  C.~Amsler {\it et al.}  [Particle Data Group],
  Phys.\ Lett.\  B {\bf 667} (2008) 1.

\bibitem{:2009ec}
  Tevatron Electroweak Working Group,
  arXiv:0903.2503 [hep-ex].

\bibitem{Nadolsky:2008zw}
  P.~M.~Nadolsky {\it et al.},
  Phys.\ Rev.\  D {\bf 78} (2008) 013004.

\bibitem{Porod:2003um}
  W.~Porod,
  Comput.\ Phys.\ Commun.\  {\bf 153} (2003) 275.


\bibitem{Khachatryan:2011tk}
  V.~Khachatryan {\it et al.}  [CMS Collaboration],
  arXiv:1101.1628 [hep-ex].


\bibitem{Muhlleitner:2003vg}
  M.~M\"uhlleitner, A.~Djouadi and Y.~Mambrini,
  Comput.\ Phys.\ Commun.\  {\bf 168}, 46 (2005).


\bibitem{AguilarSaavedra:2005pw}
  J.~A.~Aguilar-Saavedra {\it et al.},
  Eur.\ Phys.\ J.\  C {\bf 46} (2006) 43.


\bibitem{Allanach:2002nj}
  B.~C.~Allanach {\it et al.},
in {\it Proc. of the APS/DPF/DPB Summer Study on the Future of Particle Physics (Snowmass 2001) } ed. N.~Graf,
  Eur.\ Phys.\ J.\  C {\bf 25} (2002) 113.

\bibitem{Battaglia:2003ab}
  M.~Battaglia, A.~De Roeck, J.~R.~Ellis, F.~Gianotti, K.~A.~Olive and L.~Pape,
  Eur.\ Phys.\ J.\  C {\bf 33} (2004) 273.

\end{thebibliography}
\end{document}